\documentclass[aps,prl,twocolumn,floatfix,superscriptaddress,showpacs,amsfonts,amssymb,amsmath,preprintnumbers]{revtex4}
\usepackage{bm}
\usepackage{graphicx}

\renewcommand{\eqref}[1]{Eq.\ (\ref{#1})}

\newcommand{\figref}[1]{Fig.\ \ref{#1}}
\newcommand{\Figref}[1]{Fig.\ \ref{#1}}
\newcommand{\tabref}[1]{Tab.\ \ref{#1}}

\newcommand{\bea}{\begin{eqnarray}}
\newcommand{\eea}{\end{eqnarray}}
\newcommand{\lt}{\left}
\newcommand{\rt}{\right}

\newcommand{\la}{\langle}
\newcommand{\ra}{\rangle}

\newcommand{\dd}{\partial}
\newcommand{\vdel}{\boldsymbol{\nabla}}

\newcommand{\vU}{\mathbf{U}}
\newcommand{\vu}{\mathbf{u}}
\newcommand{\vf}{\mathbf{f}}
\newcommand{\vB}{\mathbf{B}}

\newcommand{\vy}{\mathbf{\hat y}}

\newcommand{\lf}{l_0}
\newcommand{\kf}{k_0}

\newcommand{\const}{\mathrm{const}}

\renewcommand{\Re}{\mathrm{Re}}
\newcommand{\Rm}{\mathrm{Rm}}

\newcommand{\Bsq}{\la B^2\ra}
\newcommand{\usq}{\la u^2\ra}
\newcommand{\urms}{u_\text{rms}}

\newcommand{\vuu}{\overline{\vu}}
\newcommand{\uux}{\overline{u}_x}
\newcommand{\oomegax}{\overline{\omega}_x}
\newcommand{\uuy}{\overline{u}_y}

\newcommand{\Brms}{B_\text{rms}}
\newcommand{\lB}{l_{\overline{B}}}
\newcommand{\vBB}{\overline{\vB}}
\newcommand{\BBx}{\overline{B}_x}
\newcommand{\BBy}{\overline{B}_y}
\newcommand{\BBz}{\overline{B}_z}

\begin{document}

\preprint{{\em Phys.\ Rev.\ Lett.}~{\bf 100}, 184501 (2008); E-print {\tt arXiv:0710.3359}}

\title{Generation of Magnetic Field by Combined Action of Turbulence and Shear}
\author{T.~A.~Yousef}
\affiliation{DAMTP, University of Cambridge, Cambridge CB3 0WA, United Kingdom}
\author{T.~Heinemann}
\affiliation{DAMTP, University of Cambridge, Cambridge CB3 0WA, United Kingdom}
\author{A.~A.~Schekochihin}
\affiliation{Plasma Physics Group, Blackett Laboratory, Imperial College, London SW7~2AZ, United Kingdom}
\author{N.~Kleeorin}
\affiliation{Department of Mechanical Engineering, 
The Ben-Gurion University of the Negev, P.~O.~Box 653, Beer-Sheva 84105, Israel}
\author{I.~Rogachevskii}
\affiliation{Department of Mechanical Engineering, 
The Ben-Gurion University of the Negev, P.~O.~Box 653, Beer-Sheva 84105, Israel}
\author{A.~B.~Iskakov}
\affiliation{Department of Physics and Astronomy, UCLA, Los Angeles, California 90095-1547, USA}
\affiliation{Plasma Physics Group, Blackett Laboratory, Imperial College, London SW7~2AZ, United Kingdom}
\affiliation{Wolfgang Pauli Institute, University of Vienna, Nordbergstra{\ss}e~15, A-1090 Vienna, Austria}
\author{S.~C.~Cowley}
\affiliation{Department of Physics and Astronomy, UCLA, Los Angeles, California 90095-1547, USA}
\affiliation{Plasma Physics Group, Blackett Laboratory, Imperial College, London SW7~2AZ, United Kingdom}
\author{J.~C.~McWilliams}
\affiliation{Department of Atmospheric Sciences, UCLA, Los Angeles, California 90095-1565, USA}
\date{\today}

\begin{abstract}
The feasibility of a mean-field dynamo in nonhelical 
turbulence with superimposed linear shear is studied numerically 
in elongated shearing boxes. Exponential growth of magnetic field 
at scales much larger than the outer scale of the turbulence is found. 
The charateristic scale of the field is $\lB\propto S^{-1/2}$ 
and growth rate is $\gamma\propto S$, where $S$ is the shearing rate. 
This newly discovered shear dynamo effect 
potentially represents a very generic mechanism for 
generating large-scale magnetic fields in 
a broad class of astrophysical systems with spatially 
coherent mean flows. 
\end{abstract}


\maketitle

\paragraph{Introduction.}
Understanding the origin of cosmic magnetism is one of the fundamental 
theoretical challenges in astrophysics. The 
turbulent motions of the plasmas that make up most astrophysical 
objects are believed to be responsible for the generation of the magnetic field. 
In particular, a generic property of the turbulence of 
conducting fluid is to amplify exponentially magnetic fluctuations
at the turbulence scales or smaller 
via the fluctuation dynamo effect \cite{Batchelor,SCTMM_stokes,SICMPY_ind}. 
A distinct problem is to explain the observed presence in most astrophysical bodies 
of magnetic fields spatially coherent 
at scales larger than the outer scale of the turbulence 
(mean fields). 
Nonhelical homogeneous isotropic turbulence 
on its own cannot give rise to a mean field. 
What are then 
the large-scale properties that must be present 
in a turbulent system for such a field to be generated? 
Mean-field dynamo theories \cite{Krause_Raedler}
have identified a number of amplification mechanisms. 
We know, e.g., that non-zero net helicity 
(often combined with rotation in real systems) is sufficient to 
produce mean fields, but is it necessary? 

Perhaps the most common large-scale feature is 
a mean velocity shear. It is present, e.g., in 
stellar interiors \cite{Ossendrijver_review}, 
accretion disks \cite{Balbus_Hawley}, 
galaxies (in particular, irregular ones \cite{Chyzy_etal}), 
and liquid-metal laboratory dynamos \cite{Monchaux_etal}, 
all of which host both large-scale (mean) 
and small-scale (fluctuating) magnetic fields. 
A number of theories 
have proposed that a mere combination of 
turbulence and shear could give rise to a mean-field dynamo:
e.g., the shear-current effect \cite{RK03}, 
the stochastic $\alpha$ effect \cite{Proctor}, 
shear amplification of small-scale-dynamo-generated field \cite{Blackman}, 
negative-diffusivity-type theories \cite{Urpin}. 
Ref.~\cite{RK03}, which used  the $\tau$-approximation closure, 
provoked a particular debate because 
it seemed to contradict the rigorous mean field theory 
based on the second-order correlation approximation (SOCA), 
which ruled out the shear dynamo 
\cite{Raedler_Stepanov}. 
However, the SOCA is only valid 
in the limit either of low hydrodynamic and magnetic Reynolds numbers, 
$\Re,\Rm\ll1$, 
or short velocity correlation times 
\cite{Krause_Raedler}. 
The real turbulent systems are in neither of these limits, 
and the hope that some of the results 
qualitatively carry over has had to be backed up by numerical evidence 
\cite{Brandenburg_alpha} and by 
intuitive physical field-amplification scenarios \cite{Parker55}. 
In the absence of a compelling physical argument 
for or against the shear dynamo or of a rigorous method for 
proceeding analytically, a numerical experiment is overdue. 
Here we report the first such numerical experiment, 
which supports the existence of the shear dynamo.

\paragraph{Numerical Set Up.}
We consider the incompressible magnetohydrodynamics 
(MHD) with a background linear shear flow 
$\vU=-Sx\vy$ and a white-noise nonhelical random 
homogeneous isotropic body force $\vf$:
\bea
\label{u_eq}
{d\vu\over dt} &=& u_xS\vy
-{\vdel p\over\rho} + {\vB\cdot\vdel\vB\over4\pi\rho} + \nu\nabla^2\vu + \vf,\\
{d\vB\over dt} &=& - B_x S\vy + \vB\cdot\vdel\vu + \eta\nabla^2\vB,
\label{B_eq}
\eea
where $\vu$ and $\vB$ are the velocity and magnetic fields, 
$d/dt = \dd_t - Sx\dd_y + \vu\cdot\vdel$, 
the density $\rho=1$, and the pressure $p$ is determined by the 
incompressibility condition $\vdel\cdot\vu=0$.  
These equations are solved with shear-periodic boundary conditions 
by a Lagrangian spectral method \cite{Lithwick_shear}. 
When the imposed shear $S$ is weak compared to the turnover
rate of the turbulent motions, the growth of the mean (large-scale)
field can only be detected if the size of the computational domain 
is much larger than the turbulence scale $\lf$. In general, this, together 
with the necessity to run the simulations for very long times, 
requires unaffordable amounts of computing power. 
We circumvent this problem by using computational boxes with 
large aspect ratios, $L_x\times L_y\times L_z$, where 
$L_z\gg L_x=L_y$. The units of length and time are fixed by setting 
$L_x=L_y=1$ and the mean forcing power $\epsilon=\la\vu\cdot\vf\ra=1$
(this can be controlled because the forcing is white-noise). 
The average forcing scale is $\lf=1/3$, i.e., the energy is injected randomly 
in the wave-number shell centered at $\kf/2\pi=3$.
The resulting root-mean-square velocity field is 
$\urms\equiv\usq^{1/2}\simeq1\pm0.3$, so the typical turnover 
rate of the turbulent motions is $\urms/\lf\sim3$.  
We study five values of the shear $S=2,1,1/2,1/4,1/8<\urms/\lf$
(weak shear is used in order for the mean and fluctuating 
fields to be clearly distinguishable from each other via 
small-scale averaging; see \eqref{BB_def} below). 
The viscosity and magnetic diffusivity are $\nu=\eta=10^{-2}$, 
so $\Rm=\Re\equiv\urms/\kf\nu\sim 5$. The resolution requirements 
are consequently not large: it suffices to have 
$32\times32$ collocation points in the $(x,y)$ plane. 
In the $z$ direction, we use resolutions between 256 
and 4096 collocation points for $L_z=8,\dots,128$, depending 
on $S$ (\tabref{tab_runs}). 

\begin{table}[t]
\caption{\label{tab_runs} Index of runs}
\begin{ruledtabular}
\begin{tabular}{llllrr}
$S$   &$L_z$&\text{Resolution}& $\gamma$&$\lB$&$\BBy/\BBx$\footnotemark[1]\\
\hline
2     & 8   &$32^2\times256$  & 0.0161 & 3.7 & 6.54      \\ 
2     & 16  &$32^2\times512$  & 0.021  & 3.8 & 6.49      \\ 
\hline
1     & 8   &$32^2\times256$  & 0.0030 & 4.6 & 6.38      \\ 
1     & 16  &$32^2\times512$  & 0.0124 & 5.4 & 6.50      \\ 
1     & 32  &$32^2\times1024$ & 0.0092 & 5.2 & 6.43      \\ 
1     & 64  &$32^2\times2048$ & 0.0121 & 5.1 & 6.35      \\ 
\hline
0.5   & 16  &$32^2\times512$  & 0.0040 & 6.8 & 6.34      \\ 
0.5   & 32  &$32^2\times1024$ & 0.0058 & 7.1 & 6.31      \\ 
0.5   & 64  &$32^2\times2048$ & 0.0055 & 7.3 & 6.32      \\ 
\hline
0.25  & 64  &$32^2\times2048$ & 0.0025 &  9.7& 6.07      \\ 
0.25  & 128 &$32^2\times4096$ & 0.0025 &  9.9& 6.06      \\ 
\hline
0.125 & 64  &$32^2\times2048$ & 0.00094& 13.1& 6.01      \\ 
0.125 & 128 &$32^2\times4096$ & 0.00092& 13.5& 6.02      \\ 
\end{tabular}
\end{ruledtabular}
\footnotetext[1]{This is the time average of 
$\lt[\int dz\BBy^2(z)/\int dz\BBx^2(z)\rt]^{1/2}$.} 
\end{table}

Strictly speaking, we cannot 
speak about turbulence with such low $\Re$. However, 
a developed inertial range is not important 
for mean field dynamos: it is sufficient 
that a stochastic velocity field with $\Re\gtrsim1$ is 
present \footnote{We found that the shear dynamo effect 
disappears when $\Re$ and $\Rm$ are too small, as predicted by 
SOCA \cite{Raedler_Stepanov}.}. 
In our simulations, $\Rm$ is subcritical with respect to 
the fluctuation dynamo \cite{SCTMM_stokes,SICMPY_ind}, 
so any field growth we detect is due 
purely to a mean-field dynamo. Note, however, that since $\Rm>1$, 
turbulent tangling of the mean field 
generates small-scale magnetic fluctuations whose energy 
is in general not smaller than that of the mean field \cite{SICMPY_ind}. 

\begin{figure}[t]
\includegraphics[width=8.5cm]{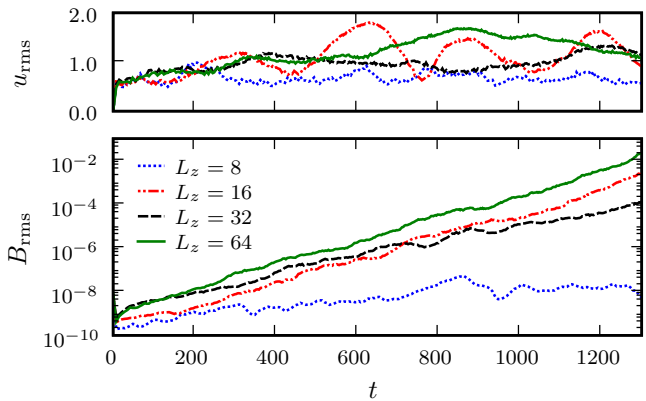}
\vskip-0.4cm
\caption{\label{fig_growth} Evolution of $\urms$ (upper panel) 
and $\Brms$ (lower panel) for $S=1$ and four values $L_z=8,16,32,64$.}
\vskip-0.3cm
\end{figure}

\paragraph{Results.}
All runs are initialized with a random, zero-mean, 
dynamically weak ($\Bsq=10^{-20}$) magnetic field. 
The field grows exponentially with time 
at all values of $S$ studied, provided the computational box is 
sufficiently long. For each $S$, we consider  
the growth rate $\gamma$ of $\Brms\equiv\Bsq^{1/2}$ to be converged 
if it stays approximately the same when $L_z$ is doubled 
(\figref{fig_growth}).
That we are able to find such values means 
that $\gamma$ is asymptotically independent 
of $L_z$ (the dependence on $L_x$ and $L_y$ should 
also be studied but that is currently too expensive computationally). 
The field eventually grows to a dynamically strong saturated level. 
Here we concentrate on the kinematic (weak-field) regime 
and leave the properties of the saturated state to a future study. 


\begin{figure}[t]
\includegraphics[width=9cm]{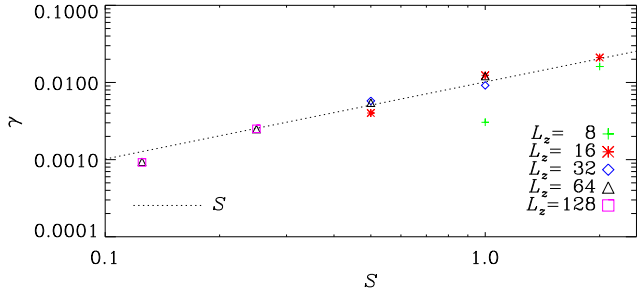} 
\vskip-0.4cm
\caption{\label{fig_gamma} Growth rates $\gamma$ of $\Brms$ for all 
runs (\tabref{tab_runs}). The dotted line shows the slope 
corresponding to $\gamma\propto S$.}
\vskip-0.3cm
\end{figure}

\Figref{fig_gamma} shows that, in the range of shears studied, 
the growth rate of $\Brms$ appears to increase linearly with $S$, 
$\gamma\propto S$. 
This is a somewhat unexpected result from the theoretical point of view
because the shear-current effect \cite{RK03}, as well as most other mean-field 
theories quoted above predict $\gamma\propto S^2$ for the fastest-growing mode.
We cannot, however, exclude the possibility that 
the $S^2$ scaling may be asymptotically recovered at even smaller~$S$.

\begin{figure}[b]
\hskip0.3cm\includegraphics[width=7.7cm]{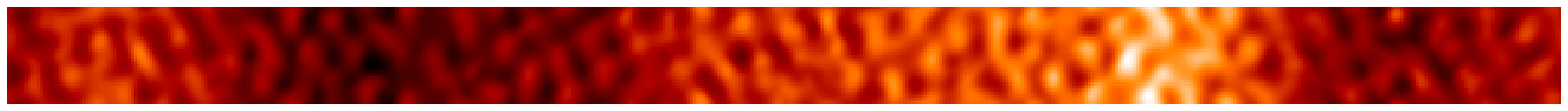}\\
\includegraphics[width=8.5cm]{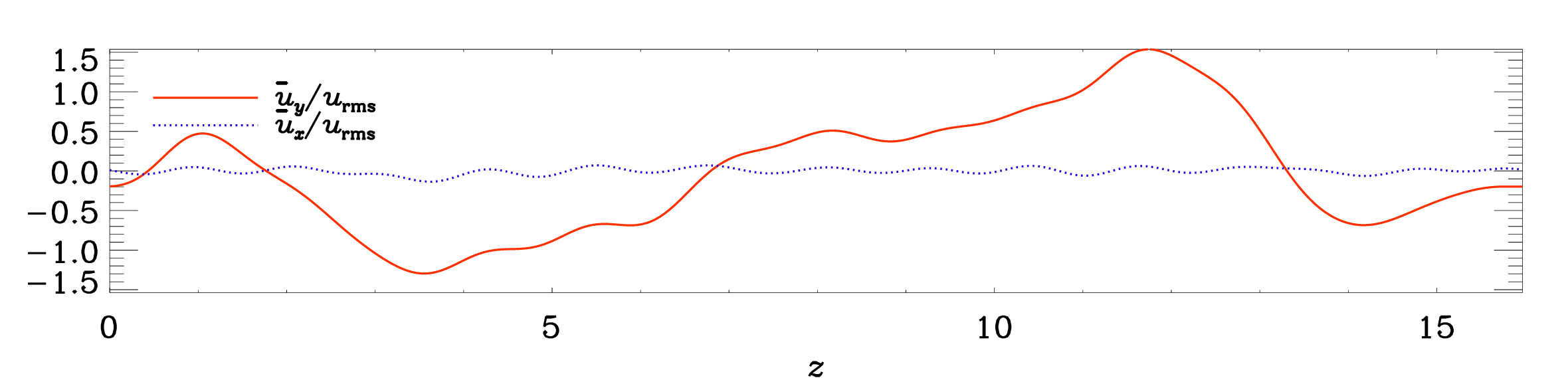}\\
\hskip0.3cm\includegraphics[width=7.7cm]{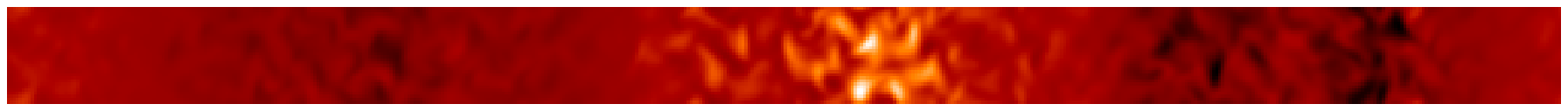}\\
\includegraphics[width=8.5cm]{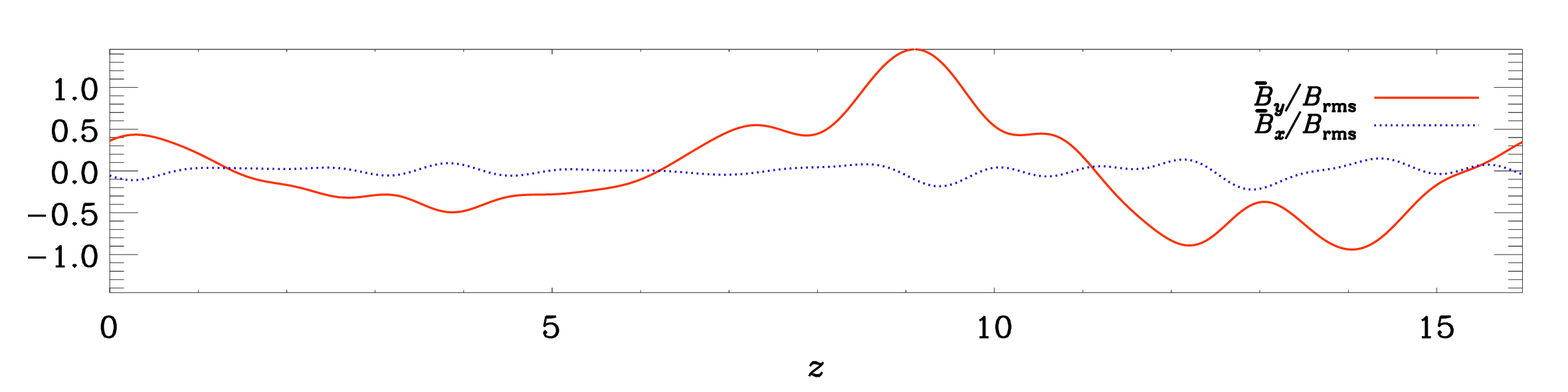}
\vskip-0.3cm
\caption{\label{fig_slices} Snapshots of $u_y$ (upper panel) 
and $B_y$ (lower panel) taken in an $(y,z)$ cross-section of the 
$L_z=16$ run for $S=1$ (the forcing scale is $\lf=1/3$). 
Underneath the snapshots are plots 
of $\uuy(z)$, $\uux(z)$ (upper panel) and 
$\BBy(z)$, $\BBx(z)$ (lower panel). Here $\vuu(z)$ is defined 
similarly to $\vBB(z)$ [\eqref{BB_def}].}
\vskip-0.3cm
\end{figure}

That the growing field is large-scale is obvious already from 
the visualization of the field: the 
large-scale $z$-dependent modulation is evident against 
the turbulence-scale structure (\figref{fig_slices}). 
We isolate this large-scale dependence 
on $z$ by low-pass filtering in Fourier space:
\bea
\label{BB_def}
\vBB(z) = \sum_{|k_z/2\pi|<1}\vB(k_x=0,k_y=0,k_z)\,e^{ik_z z}.
\eea 
Note that since $\vdel\cdot\vBB=0$, $\BBz=0$. 
This procedure averages out the small-scale 
structure and brings out the growing large-scale field in 
a clear way (\figref{fig_slices}). 
In all cases, the root-mean-square values of $\BBx$ 
and $\BBy$ grow exponentially with the same rate $\gamma$ as $\Brms$. 
We find $|\BBy|>|\BBx|$, which is expected because the shear 
systematically converts $B_x$ into $B_y$ [\eqref{B_eq}]. 
The ratio $|\BBy/\BBx|\sim S/\gamma$ is approximately constant in time 
and its average is independent of $S$ (\tabref{tab_runs}), 
which is consistent with $\gamma\propto S$ established above.

\begin{figure}[t]
\includegraphics[width=9cm]{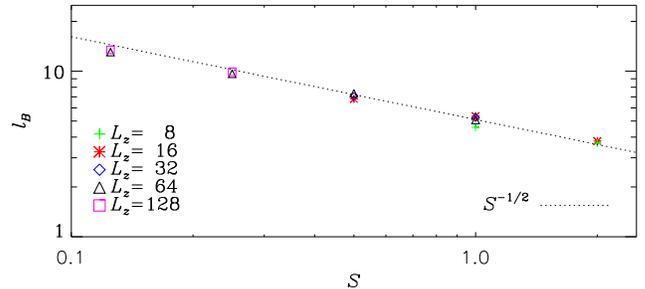} 
\vskip-0.4cm
\caption{\label{fig_scales} 
The characteristic scale of the magnetic field [\eqref{lB_def}] for all runs. 
The dotted line showes the slope $S^{-1/2}$.}
\vskip-0.3cm
\end{figure}

Examining \figref{fig_slices}, 
we see that the magnetic field grows in large random patches. 
In time, they move around (in $z$) and change shape 
in an apparently random fashion. 
This means that the mean field is not 
strictly speaking an ``eigenmode'' with a spatial profile  
constant in time, although we found the correlation time scale of 
its evolution to be much longer than 
either the turnover time ($\lf/\urms$) or the shear time ($S^{-1}$). 
In order to describe the spatial structure of the mean field 
in a systematic way, we define the time-averaged characteristic scale~$\lB$: 
\bea
\label{lB_def}
{1\over\lB} = {1\over t_2-t_1}\int_{t_1}^{t_2} dt 
\lt[{\int dz\lt(\dd\BBy/\dd z\rt)^2\over\int dz \BBy^2}\rt]^{1/2}.
\eea
Here and in all other cases, 
the time average is taken over the exponential-growth 
(kinematic) period $(t_1,t_2)$ of the field evolution. 
The derivatives are calculated in Fourier space.  
The values of $\lB$ are given in \tabref{tab_runs} and 
plotted vs.\ $S$ in \figref{fig_scales}. As the shear is 
decreased, $\lB$ increases and is well 
matched by the scaling $\lB\propto S^{-1/2}$. 

\begin{figure}[b]
\includegraphics[width=8.5cm]{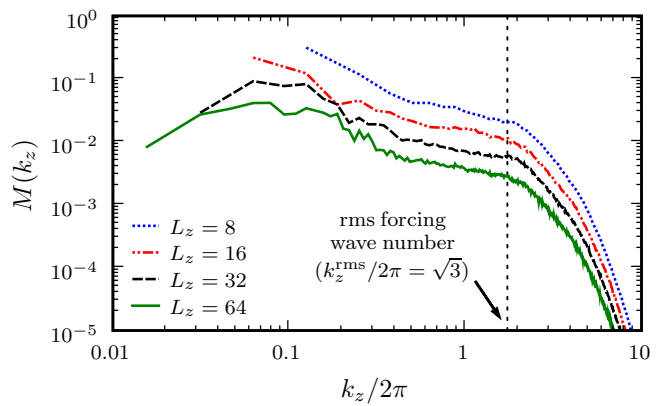}
\vskip-0.3cm
\caption{\label{fig_spectra} Normalized one-dimensional spectra 
of the magnetic energy, 
$M(k_z)=\sum_{k_x,k_y} |\vB(k_x,k_y,k_z)|^2/\Bsq$ (time averaged 
over the growth stage) for $S=1$ and $L_z=8,16,32,64$. 
The four graphs demonstrate that, as $L_z$ is increased, 
a large-scale spatial structure independent of the box length 
emerges.} 
\vskip-0.3cm
\end{figure}

This scaling is again at odds with the mean-field-theory prediction 
$\lB\propto S^{-1}$ \cite{RK03}, but the following simple 
argument shows that it is consistent with 
$\gamma\propto S$ and suggests the 
possible form that a mean-field theory of the dynamo reported 
here may take. 
Let us conjecture the mean-field 
equations in the following (standard) model form
\bea
\dd_t\BBx &=& -\eta_T k_z^2\BBx - A\BBy,\\
\dd_t\BBy &=& -\eta_T k_z^2\BBy - S\BBx,
\eea
where $\eta_T\sim\urms\lf$ is the turbulent diffusivity 
and $A$ is some operator that closes the dynamo loop.
(the challenge of mean-field theory is to find $A$). 
The growth rate is $\gamma=\sqrt{SA}-\eta_T k_z^2$. 
For the fastest-growing mode, 
the two terms in this expression are comparable, 
so, if $k_z\sim\lB^{-1}\sim\lf^{-1}(S\lf/\urms)^{1/2}$, then 
$A\sim S$ and $\gamma\sim S$ (in contrast, some mean-field theories predict 
$A\sim S \lf^2 k_z^2$ \cite{RK03,Proctor}, so $k_z\propto S$ and $\gamma\propto S^2$). 

Finally, in \figref{fig_spectra}, we show the 
spectrum of magnetic energy during the growth stage.
It is strongly peaked at large scales ($k_z\lf\ll1$), 
but also shows that the mean field is tangled 
by the turbulence to produce a significant amount 
of magnetic energy at the turbulence scales. 
The presence of this tangled component is likely to be 
essential in the shear dynamo effect in that the 
small-scale field could be continuously resupplying 
the field $B_x$ for amplification by shear \cite{Blackman}.
The key to a successful theory of the shear dynamo may be to find 
the way in which the symmetry is broken to make this resupply 
systematically favorable to the exponential growth of the mean 
field (as, e.g., implicitly attempted in Ref.~\cite{RK03}). 

\paragraph{Effect of Shear on Velocity Field.} 
It has been suggested that 
a mechanism similar to the shear dynamo 
may also produce large-scale velocity structures 
(``vorticity dynamo'' \cite{EKR03}). 
The velocity does indeed develop 
large fluctuations that are energetically comparable 
to the small-scale turbulence, 
last for long times (\figref{fig_growth}) and 
are spatially coherent on scales similar 
to those of the magnetic field (\figref{fig_slices}). 
The large-scale structure forms 
mainly in $u_y$ (corresponding to 
large-scale vorticity, $\oomegax=-\dd\uuy/\dd z$). 
As the forcing always seeds some large-scale vorticity 
that is not infinitesimally small, the ``vorticity dynamo'' 
always operates in the nonlinear regime. 
Its detailed study is outside the scope of this Letter. 

The growth of the magnetic field does not seem to be strongly correlated 
with the evolution of the shear-generated large-scale velocity structures 
(compare, e.g., the time evolution of $\urms$ and $\Brms$
in \figref{fig_growth}). 

It is well known that the presence of shear can lead 
to nonlinear destabilization of finite velocity fluctuations
and formation of shear-driven turbulence 
whose outer scale is the scale of the shear (in simulations 
with a linear shear, the box scale). 
This does indeed happen in our simulations when $S$ is too strong 
or the box is too long. The quantitative signature 
of this regime is that the power input from the shear 
in \eqref{u_eq}, $\la u_x u_y\ra S$, exceeds the 
forcing power $\epsilon=\la\vu\cdot\vf\ra$. We avoid this regime 
to isolate the mean-field generation effect, which requires 
a scale separation between the 
turbulence and the mean field. In all runs reported here, 
$|\la u_x u_y\ra S|\ll\epsilon$. We note that the large 
upward fluctuations of $\urms$ [\figref{fig_growth}] 
are not accompanied by a significant change 
in $\la u_x u_y\ra S$, so the large-scale 
velocity structures appear to feed on the forcing power, 
not on the power extracted from the shear.  

\paragraph{Discussion.} 
We have found that a large-scale magnetic field grows 
exponentially in long sheared boxes with forced small-scale 
nonhelical turbulence. In the parameter range we have studied, 
the growth rate appears to scale as $\gamma\propto S$, the 
spatial scale of the field $\lB\propto S^{-1/2}$ and $\BBy/\BBx\simeq\const>1$. 
These properties do not seem to fit any of the existing theories. 
Our results do, however, lend credence to the concept 
of a shear dynamo and thus provide motivation for 
further theoretical effort. 

To our knowledge, this is the first demonstration of 
the shear dynamo effect in a dedicated numerical experiment. 
In an earlier unpublished study 
we obtained similar results 
using PENCIL code (a compressible finite-difference code in contrast to 
the spectral one used above), so 
the amplification effect appears to be numerically robust. 
We note that there have been earlier 
indications of nonhelical turbulence amplifying 
large-scale magnetic field in the presence of a large-scale 
shear associated with mean flows in numerical experiments 
that used constant-in-time sinusoidal forcing functions 
\cite{Brandenburg_shear,Shapovalov}. 
Another example of large-scale magnetic fields generated by 
a combination of nonhelical turbulence and a mean flow is 
the numerical experiments with Taylor-Green forcing 
\cite{Ponty_etal1}. 
One might speculate that the shear provided 
by the mean flow in such systems could act in a way qualitatively 
similar to a linear shear and give rise to 
mean-field amplification. 

Another context in which 
the shear dynamo may be important is 
accretion disks, where turbulence is driven by 
the magnetorotational instability (MRI) \cite{Balbus_Hawley}. 
The MRI requires a (weak) large-scale field and gives 
rise to velocity and magnetic fluctuations at small scales. 
This turbulence could then conceivably couple to the 
large-scale Keplerian shear and amplify the large-scale field, 
thus closing the loop. While the feasibility of such an MRI-dynamo 
mechanism will be the object of a separate study, 
we have verified already that the shear dynamo 
continues to work in the presence of rotation. 


\begin{acknowledgments} 
We thank A.~Brandenburg, L.~Kitchatinov, G.~Lesur, G.~Ogilvie, J.~Papaloizou, 
D.~Shapovalov, D.~Sokoloff and especially F.~Rincon for valuable discussions. 
Simulations were done on UKAFF, NCSA (Illinois) and Cambridge HPCF. 
This work was supported by STFC (T.H., A.A.S),  
Newton Trust (T.A.Y, T.H.), 
US DOE CMPD (A.B.I.),
the Royal Society (I.R., A.B.I.)
and the Leverhulme Trust Network for 
Magnetized Plasma Turbulence. 
\end{acknowledgments}


\end{document}